\documentclass[usenatbib]{mn2e}

\input{epsf}
\newcommand{\be}{\begin{eqnarray}}
\newcommand{\ee}{\end{eqnarray}}
\newcommand{\hMpc}{{\ifmmode{h^{-1}{\rm Mpc}}
\else{$h^{-1}$Mpc}\fi}}
\newcommand{\hkpc}{{\ifmmode{h^{-1}{\rm kpc}}
\else{$h^{-1}$kpc}\fi}}
\newcommand{\hM}{{\ifmmode{$h^{-1}{\rm M}_\odot$}
\else{$h^{-1}{\rm M}_\odot$}\fi}}

\title[Multicomponent DM model: haloes]
{Dark matter halo formation in the multicomponent dark matter models}
\author[Semenov, Pilipenko, Doroshkevich, Lukash, Mikheeva]
{V.A. Semenov$^{1,2}$, S.V. Pilipenko$^{1,2}$, A.G. Doroshkevich$^1$, V.N. Lukash$^1$, \newauthor E.V. Mikheeva$^1$\\
$^1$ Astro Space Center of Lebedev Physical
           Institute of  Russian Academy of Sciences, Profsojuznaja st. 84/32,
                        117997 Moscow,  Russia\\
        $^2$ Moscow Institute of Physics and Technology, Institutskij per. 9, 
            141700 Dolgoprudnyj, Russia\\}
\date{Accepted ....,
      Received ...,
        in original form ... .}

\begin{document}
\maketitle
\begin{abstract}
This work investigates a set of cosmological collisionless N-body simulations with featured power spectra of initial perturbations in the context of the core-cusp and satellite problems. On the studied power spectra some scales of fluctuations were suppressed. Such spectral features can be caused by multicomponent dark matter. The density profiles innermost resolved slopes $\alpha\equiv d \log(\rho) /d \log(r) $ of the five largest haloes were measured and its dependence on the parameters of the suppression was traced. In a certain range of the parameters the slopes flatten from initial value of about $-1.2$ to $-0.6$ or even to $-0.2$ in one of the cases. This qualitatively demonstrates that (i) profiles shape depends on initial power spectrum and (ii) this effect may be responsible for the solution of the core-cusp problem. The suppression of some part of the initial power spectrum also leads to the decrease of the number of massive subhaloes.

\end{abstract}

\section{Introduction}
The so-called ``core-cusp problem'' is one of the essential problems in the standard $\Lambda$CDM cosmology. The problem lies in discrepancy of dark matter haloes inner structure in observations and in numerical simulations. It appeared that the central slopes of density profiles $\rho(r)$ in dwarf and LSB-galaxies is significantly smaller than those obtained in numerical simulations (e.g. \citealt{Burkert1995,kravtsov1998,salucci2000,deblok,deBlok2010}). However, cuspy haloes with steep central slope $\alpha>1$ are observed as well, yet in massive galactic clusters \citep{2003ApJ...586..135L,2002ApJ...572...66A}. Measurement of density profiles in spiral and elliptical galaxies is aggravated by total domination of bright baryon matter in their central parts. Thus, cuspless profiles are observed only in a special type of galaxies dominated by dark matter.

Recently new evidence appeared on that galactic cusps in some cases can be destructed by the firm AGN feedback \citep{2013MNRAS.tmp.1290M}. Apart from this, wide variety of exotic non-standard models of dark matter were proposed: warm \citep{wdm0,wdm}, collisional \citep{sidm}, fluid \citep{fluiddm}, decaying \citep{decdm0,decdm1,2009ARep...53..976P}, flavour-mixed or oscillating \citep{2012APS..APR.G7007M}. 

\citet{DoLuMa} claimed that the core-cusp problem can be solved within the standard model. Galactic cusps can be eliminated by small-scale motions of particles which are not accounted for in numerical simulation. Particles motions may be characterised by the coarse-grained entropy function $E$:
\[
E = \sigma^2 \rho^{-2/3},
\]
where $\sigma$ is the velocity dispersion of dark matter particles and $\rho$ is their density. As was shown in \citet{DoLuMa}, $E$ can be increased by either initial temperature of particles or by small scale density perturbations. Regardless the source of entropy, the growth of $E$ should result in the flattening of density profiles since cuspy profile has $E\rightarrow 0$ when $r\rightarrow 0$.

Numerical investigation of this effect in toy models of halo formation by \citet{PDLM:2012} confirmed the flattening of cusp by small scale perturbations. The perturbations on the scale $k\sim 10k_L$ to $k\sim 20k_L$, where $k_L$ stands for the main large-scale frequency of the halo, are responsible for this effect. However also the importance of perturbations of \textit{intermediate} scales with $k<10k_L$ was demonstrated. These perturbations retain cusp if their amplitude is high enough. Thus, in order to remove cusps the intermediate scale perturbations should be damped. \citet{PDLM:2012} argued that this can happen quite rare for the standard CDM power spectrum: the probability of having intermediate perturbations damped for a given halo is $\leq1$\%.

In this paper we further investigate the interaction of perturbations of different scales in more realistic models with initial power spectrum constructed from the standard one by the introduction of a spectral feature.  In this study particular types of spectrum features were chosen to qualitatively reproduce initial conditions used by \citet{PDLM:2012}.

We suggest a physical model of how this feature can be produced in the Universe. The spectral feature should appear if the dark matter consists of at least two species with different particle mass. Some interaction between these species can be responsible for the complex shape of power spectrum. In particular, very interesting model is anapole dark matter \citep{anapole}. Thus, the core-cusp problem and galactic satellites problem of the $\Lambda$CDM cosmology may give us some insights into the physics of dark matter.

The effect of featured power spectra on large-scale structure was previously studied by \citet{2001MNRAS.326..109K}. Authors added bump and various dips on standard power spectrum and found that this changes only halo-halo correlation function and haloes peculiar velocities (in case of the bumpy spectrum). According to the authors, local halo abundance and large-scale velocities field remain unchanged. In regard to density profiles, although the authors found no difference between the models, actual density profiles provided in the paper shows slight flattening of cusps for all modified models. Absence or weakness of the effect may be explained by the insufficient resolution of the models and the chosen parameters of the spectrum dips. According to \citet{PDLM:2012} the dip should be broad enough to produce the effect. Only the largest dip in \citep{2001MNRAS.326..109K} meets this condition. Additionally, dip's shape was chosen so that it did not suppress all intermediate frequencies. 

This paper structured as follows. Section 2 describes parameters of numerical simulations and implemented features of power spectra. In section 3 we discuss main results and their connections to the entropy theory. Section 4 concludes.

\begin{figure*}
\centering
\epsfxsize=\textwidth
\epsfbox{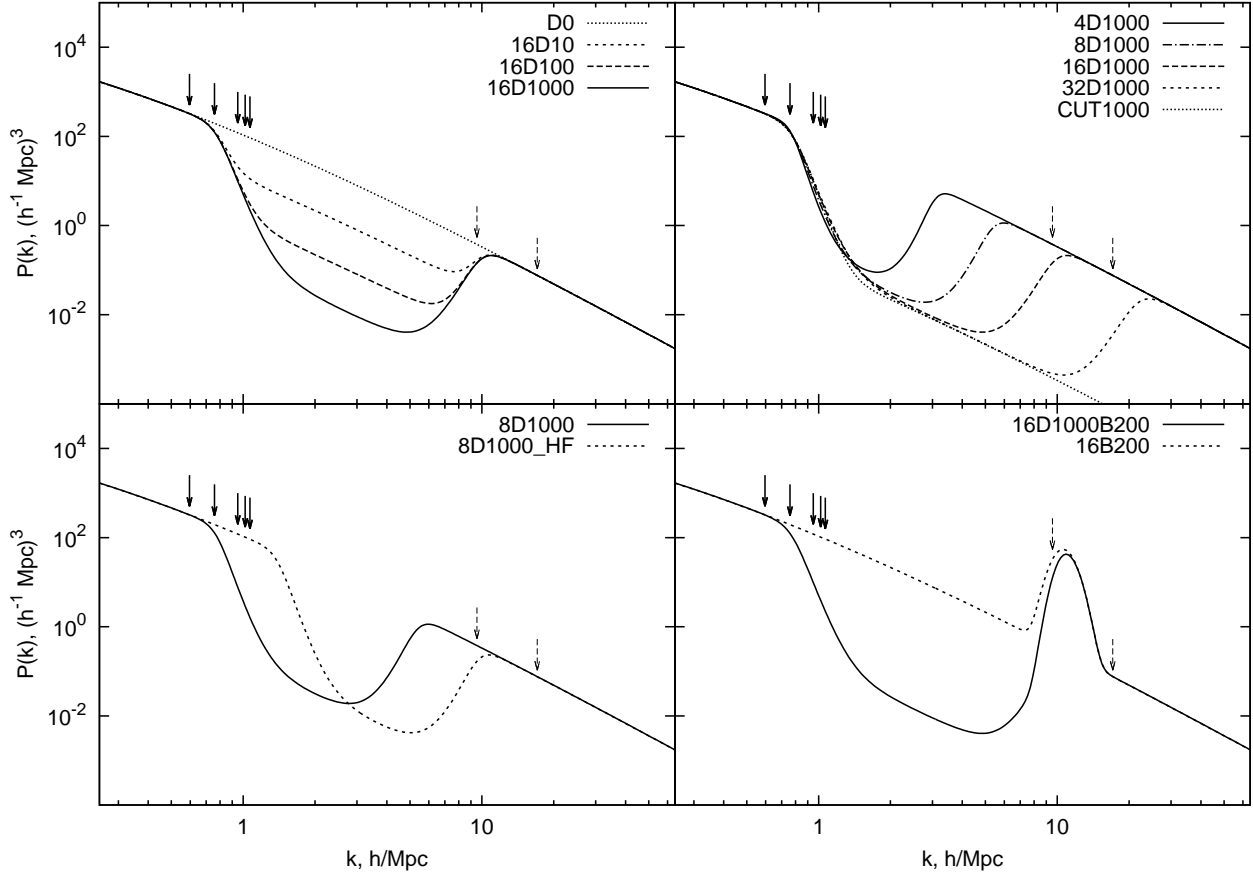}
\caption{Modified power spectra. \textit{Top left:} dips with various depths. \textit{Top right:} dips with various widths. \textit{Bottom left:} two dips with different positions. \textit{Bottom right:} combination of features of different types (16D1000B200) and bumpy spectrum (16B200). The solid arrows show scales of the largest haloes in the simulation (H1--H5), left and right dashed arrows indicate scales 16 times smaller than that of H1 and H5 respectively.}
\label{fig:models}
\end{figure*}

\section{N-Body Simulations with Modified Power Spectra}

To study the effect of various spectrum features on properties of the inner structure of haloes the numerical simulations were run in a cubical volume 25 $h^{-1}$Mpc in a side with periodic boundary conditions. $256^3$ probe dark matter particles were used with corresponding mass resolution $7.76\cdot 10^7$ $h^{-1}{\rm M}_\odot$. The comoving force softening length was chosen to be $\min(L_i/5.5,L_i/55a)$, where $L_i$ is a mean initial separation between particles which is $\sim$0.1 \hMpc. The simulations were performed with the aid of GADGET-2 code \citep{gadget2}. The concordance $\Lambda$CDM parameters were used (the dimensionless density parameters $\Omega_{\rm m}=0.3$ and $\Omega_\Lambda=0.7$). Initial conditions were set using N-GenIC code written by Volker Springel.

Along with the standard $\Lambda$CDM power spectrum several modified on small-scales spectra were studied. Features were introduced as:
\[
P(k)=F(k)\cdot P_0(k),
\]
where $P_0(k)$ is the standard spectrum and $F(k)$ defines actual shape of the feature. In case of bumpy spectra used $F(k)$ was similar to one chosen by \citet{2001MNRAS.326..109K}:
\[
F(k)=F_{b}(k,k_b,\sigma_b,A_b), \nonumber \\
\]
\[
F_{b}(k,k_b,\sigma_b,A_b) \equiv 1 + A_b \cdot\exp{\left(-\frac{\log^2(k/k_{b})}{2\sigma_{b}^2}\right)},
\]
which introduces logarithmic Gaussian bump on scale $k_b$ with width $\sigma_b$ and amplitude $A_b$. When the amplitude equals to $A_b$, it implies that at maximum of the bump the power of perturbations $P_0(k_b)$ will be enhanced by $(1+A_b)$ times.

Another type of studied features was dip which suppressed all frequencies in range between wave numbers $k_1$ and $k_2$:
\[
F(k)=F_{d}(k,k_1,k_2,A_d,T_d), \nonumber \\
\]
\[
F_{d}(k,k_1,k_2,A_d,T_d) \equiv 1 - \frac{A_d}{ \exp\left( \frac { \log(k/k_1) \cdot \log(k/k_2)}{T_{d}} \right) + 1 },
\]
where $A_d$ defines the rate of suppressing and $T_d$ characterize the sharpness of dips boundaries. This kind of feature is of the principle interest since it qualitatively represents initial conditions used by \citet{PDLM:2012} (two delta-shaped peaks on small- and large-scales without intermediate modes). 

Finally, combination of different features was investigated. In this case small-scale perturbations are enhanced with intermediate frequencies suppressed and $F(k)$ is expressed as:
\[
F(k)=F_{b}(k,k_b,\sigma_b,A_b) \cdot F_{d}(k,k_1,k_2,A_d,T_d).
\]

Initial conditions are set in such a manner, that all corresponding spatial frequencies in different simulations have equal phases and their amplitudes are counted according to the power spectrum. In this case as long as the large-scale part of power spectra is kept unchanged, the large-scale structure (including positions and masses of the largest haloes) remains roughly the same in all simulations examined. This enables us to trace changes of properties for the same haloes in cases of different power spectra. In different simulations haloes were matched if their centers are not displaced further than their virial radii. Virial radii do not change more than 10\% for the same haloes in different simulations. The accuracy of this procedure was verified by the fact that all matched haloes in featured models consist mostly of the particles which form these haloes in the model with standard spectrum (in the two worst cases the fractions were 51\% and 65\%, in other cases it was about 80\% and even more). However, one can use this criteria only to identify the largest haloes which scales are not affected by spectral features. The five largest haloes were chosen for further investigation. In case of standard power spectrum their masses cover a range between $1.7\cdot 10^{13}$ \hM~and $9.9\cdot 10^{13}$ \hM. Masses and concentrations ($c\equiv R_{vir}/r_s$) are listed at the top of tab.~\ref{tab:results} (the haloes are numbered sequentially  according to their masses). Thus, the haloes contain sufficient number of particles ($2\div13\cdot10^5$) to study density profiles on the smallest scale of about few kpc. 

Since chosen box size and mass resolution encompass only the high-frequency monotonic part of the $\Lambda$CDM power spectrum, all obtained results are almost scale free. In other words, the results depends only on relative scales of certain haloes and spectrum features used. For this reason it is convenient to express all features scales in terms of the largest haloes' average scale $k_h \approx 0.8$ h/kpc. 

In order to study the effect of perturbations of different scales on density profiles the aforementioned features with different parameters were examined. In particular, amplitude, width and position were varied. The shapes of chosen features are shown in fig.~\ref{fig:models}. Left boundary of the dips was chosen to be $k_1=k_h$. In the top left panel the models with dips of various depth are shown. In this set of models right boundaries of dips were set on $k_2=16k_1$ scale. Thus, intermediate spatial frequencies are suppressed on scales between $k_h$ and $16k_h$ which is sufficient condition for small-scale perturbations to eliminate the density cusp\footnote{It should be kept in mind that according to \citet{PDLM:2012} for the effect of small-scale perturbations being pronounced the small fluctuations scale $k_S$ should be within a range $10k_L < k_S < 20k_L$, where $k_L$ is the scale of main collapsing mode which in this case equals to the scale of the halo $k_h$}. Dips' amplitudes $A_d$ were chosen to be 0.9, 0.99 and 0.999 for models 16D10, 16D100 and 16D1000 where intermediate modes are suppressed by a factor of 10, 100 and 1000 respectively. Model D0 represents standard power spectrum without any features. Another parameter to study was dip's width. In the top right panel the models with varied width are shown (model 4D1000 with $A_d=0.999$ and $k_2=4k_1$, model 8D1000 with $k_2=8k_1$, model 16D1000 with $k_2=16k_1$, model 32D1000 with $k_2=32k_1$ and model CUT1000 with $k_2$ much larger than $k_1$ so that all modes with scales smaller than $k_1$ were suppressed). The bottom left panel represents two models with dips of the same width ($k_2=8k_1$) yet on different scales ($k_1=k_h$ in model 8D1000 and $k_1=2k_h$ in model 8D1000\_HF). Finally, in the bottom right panel models D0 and 16D1000 are shown with introduced bump on scale $k_b=16.2k_h$.

\begin{figure}
\centering
%\epsfbox{img/profs1.eps} \\
\epsfbox{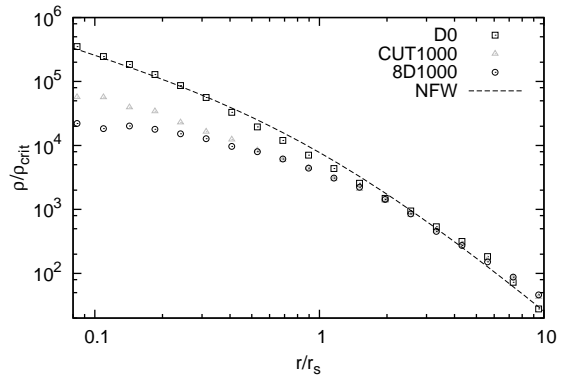} 
%\caption{\textit{top:} Density profiles of the largest halo (H1) in the models with the standard power spectrum (D0), with 1000 times suppressed intermediate modes (16D1000) and with suppressed intermediate modes and enhanced small-scale fluctuations (16D1000B200). \textit{bottom:} The most vigorous effect of spectrum modification on density profile (H4).}
\caption{The most vigorous effect of spectrum modification on density profile (H4). In model CUT1000 cusp reproduces, yet with smaller central density. This happens because of the partial suppression of the main large-scale mode.}
\label{fig:profs}
\end{figure}

\section{Results and discussion}

\subsection{Central slopes}

The largest haloes in the simulation contain about $10^5\div 10^6$ particles. Their density profile was fitted with NFW profile (see example in fig.~\ref{fig:profs}). The density profiles were found to be stable as they do not change their shape during at least five last snapshots. However, since NFW profile slope $\alpha \equiv d \log(\rho) /d \log(r) $ on small radii slowly approaches its asymptotic value $\alpha=-1$, the central slopes (measured over the innermost five bins where it remains almost the same) of the largest haloes proved to be $\alpha=-1.2$. For haloes of lower masses the situation is even less encouraging due to poorer resolution.

It is worth noting that apart from universal NFW profiles many other fits were introduced in literature, the most part of which endeavours to approximate density behaviour both in the inner and outer parts of haloes. However, it seems reasonable to describe the central part of halo separately from its outskirts, since after halo formation, its center can hardly be changed due to high speed of particles which come through the deep potential well. At the same time the outer parts of the halo continue to evolve due to mergers and mass accretion along filaments. Therefore, approximating of both regions simultaneously is not quite adequate. This problem might be solved, for instance, using the ($\alpha$, $\beta$, $\gamma$) model \citep{zhao1996,kravtsov1998} which distinguishes between these parts in a certain sense. This work examines only the innermost slope as density behaviour in the outskirts of haloes remains almost unchanged in the studied models.

The resulting inner slopes are listed in tab.~\ref{tab:results}. Different haloes have divers masses and were formed from perturbations of different scales. Therefore, their inner structure responds on spectral features in different ways. Besides, it should be kept in mind that correspondence between Fourier-space (where power spectrum is defined) and real-space (where haloes form) is not bijective. Actual shapes and structures of haloes depend on the local amplitudes of initial density perturbations which are statistically bond with power spectrum. Therefore, the relation between haloes inner structure and the shape of power spectrum should be treated statistically as well. Thus, certain trends in the density profiles shape changing can be traced.

First of all, dependence of the inner slopes on the dip's depth was examined. As it can be seen from tab.~\ref{tab:results} (models D0, 16D10, 16D100 and 16D1000) cusps of all studied haloes become shallower with higher rate of intermediate modes suppression. This result is in line with that of \citet{PDLM:2012}, where the effect of small-scale perturbations is eliminated in case if intermediate modes are sufficiently intense. The width of the dips was fixed at approximately the same value as the separation between small- and large-scale perturbations which produces the most pronounced effect in \citet{PDLM:2012}. 

On the contrary, dependence on the dip's width does not exhibit such consonance. The effect appears for different haloes on divers values of the width. %According to \citep{PDLM:2012}, the effect must not be significant in case if small-scale spectral part of perturbations is closer than 10 times of the scale of large perturbations or 
Specifically, haloes H1 and H4 have the smallest central slope when the small-scale edge of the dip is 8 times of the large-scale one (8D1000). For H2 the effect is mostly pronounced in models 8D1000 and 16D1000. In case of H5 one can observe the same (within the error) minimal value of inner slope in 16D1000 and 32D1000, while halo H3 shows shallowing of the cusp in a range of dip width (8D1000, 16D1000, 32D1000). The most vigorous effect of flattening was obtained for the halo H4 in model 8D1000, where the innermost slope proved to be $\alpha\approx -0.2$ (fig.~\ref{fig:profs}). This difference in behaviour occurs partly because of local statistical realisation of the spectra, partly because of haloes masses disparity. As it was mentioned before, haloes of different masses effectively ``feel'' the same dip like if it were placed on different relative scales. Nonetheless, with all small-scale modes suppressed in CUT1000 the inner slopes are steepen for all haloes (see also fig.~\ref{fig:profs}). This indicates that it is the small-scale perturbations which are responsible for the cusp destruction. Additionally, cusp partially (in case of H1) or entirely (in all other cases) reproduces when the dip is too narrow (4D1000). In this model intermediate modes are not suppressed on sufficient scales which results in steepening of the slope.

%As far as bumpy spectra are concerned, 
When it comes to bumpy spectra, the collapse of the bump introduces abundance of small minihaloes with masses about mass scale of the bump. It is clear from tab.~\ref{tab:results} that pure enhancing of small-scale perturbations (model 16B200) does not produce the effect. The density profiles in this case are almost indistinguishable from those obtained in the model with standard spectrum. This illustrates significance of intermediate modes which can eliminate the effect of small-scale perturbations during non-linear evolution and restore the cusp. Put simply, intermediate modes collapse before the main large-scale mode of a certain halo and resulting subhaloes fall into the center of the halo during hierarchical clustering which results in steepening the cusp. Collapsed small-scale perturbations have smaller masses and corresponding subhaloes fall longer since the falling time proportional to the ratio of the halo and satellite masses. 

Another effect caused by minihaloes is destruction of the cusp due to dynamical friction. Even though this impact is quite feeble, slow subsequent flattening of the inner slope was observed for some haloes over a several last snapshots. It should be noted here that precise measurement of central slopes is aggravated by the presence of these minihaloes since their motions induce oscillations of the density profiles.

\subsection{Central mass deficiency}

Apart from the inner slopes of density profiles the important property of haloes inner structure is the central density, or in other words the mass encompassed in some small enough region of fixed size. The central mass depends mostly on the moment of halo formation since it imprints average density of the Universe at the turnaround moment. In this work the mass shortage in models with featured spectra is characterized as the relative drop of the mass within the sphere of radius $r_s$ ($M(r_s)$) with respect to that in the model with the standard spectrum $M_{\rm D0}(r_s)$, i.e. $(M_{\rm D0}(r_s)-M(r_s))/M_{\rm D0}(r_s)$. It is evident from tab.~\ref{tab:results} that in cases when halo scale is close enough (or even lies within) to the dip on the spectrum, central mass drops. Moreover, the effect is more and more pronounced with smaller masses of haloes. Again, haloes of different masses ``feel'' the deep in different ways. While the largest haloes ``lie'' at certain distance from the dip's edge, the smallest ones are within the dip. Therefore, the main perturbations which form the haloes are suppressed to different degrees. We suppose that the main mode suppression results in latter time of formation and therefore in less central density. One can see this difference in fig.~\ref{fig:profs} for models D0 and CUT1000: even though in both cases the halo has cuspy profiles, central density is smaller in case of small-scale modes suppression. Due to latter formation time density profiles may lose stability. However, it was found that they are stable as well as corresponding profiles in standard model, yet on shorter times.

The only models with relatively weak effect of central mass deficiency are 8D1000\_HF and 16B200, where all haloes main scales are not affected by spectrum features. Likewise, drop of central mass in 16D10 is not as significant as in other models due to the mild rate of suppression.

\subsection{Subhaloes abundance}

\begin{figure}
\centering
\epsfxsize=0.49\columnwidth
\epsfbox{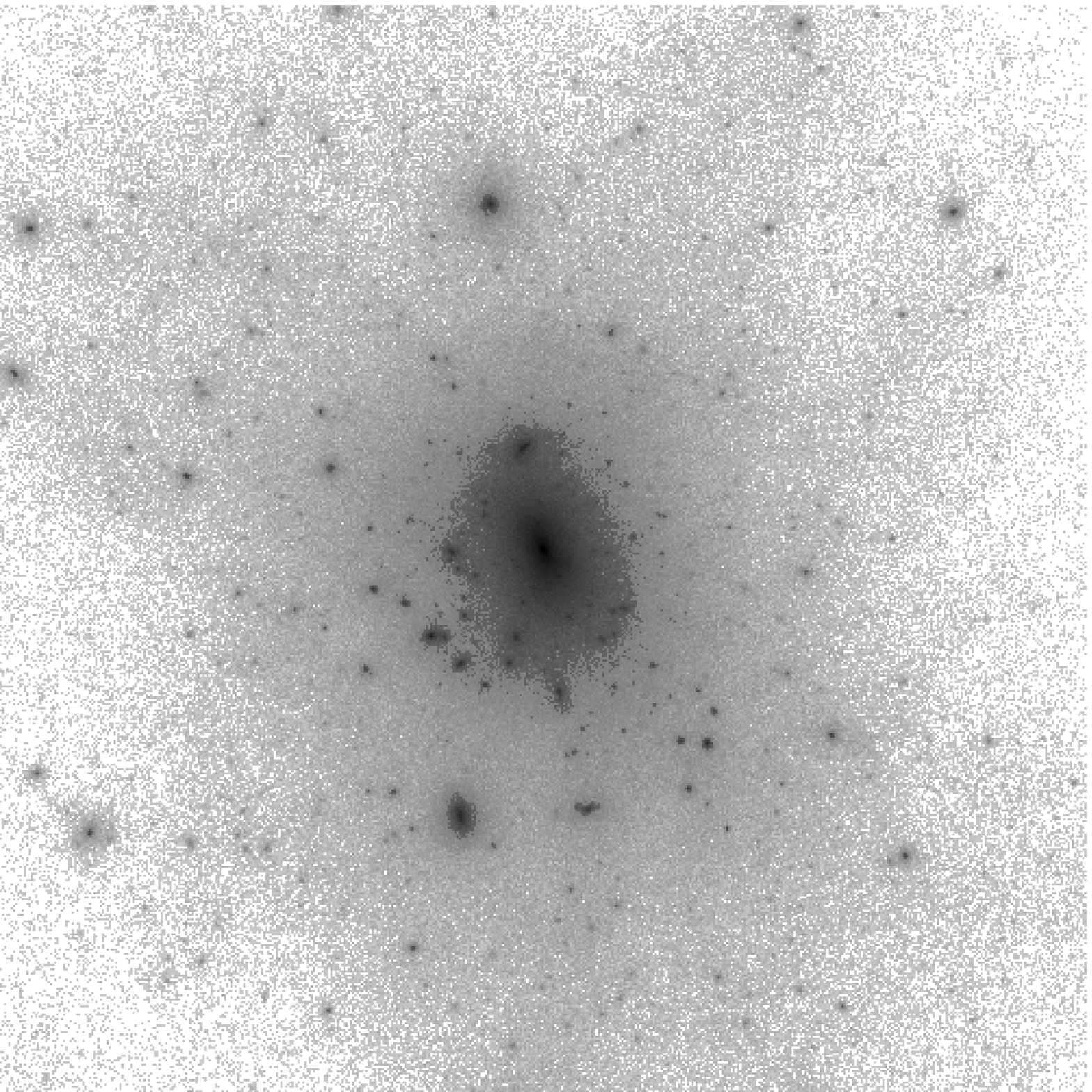} 
\epsfxsize=0.49\columnwidth
\epsfbox{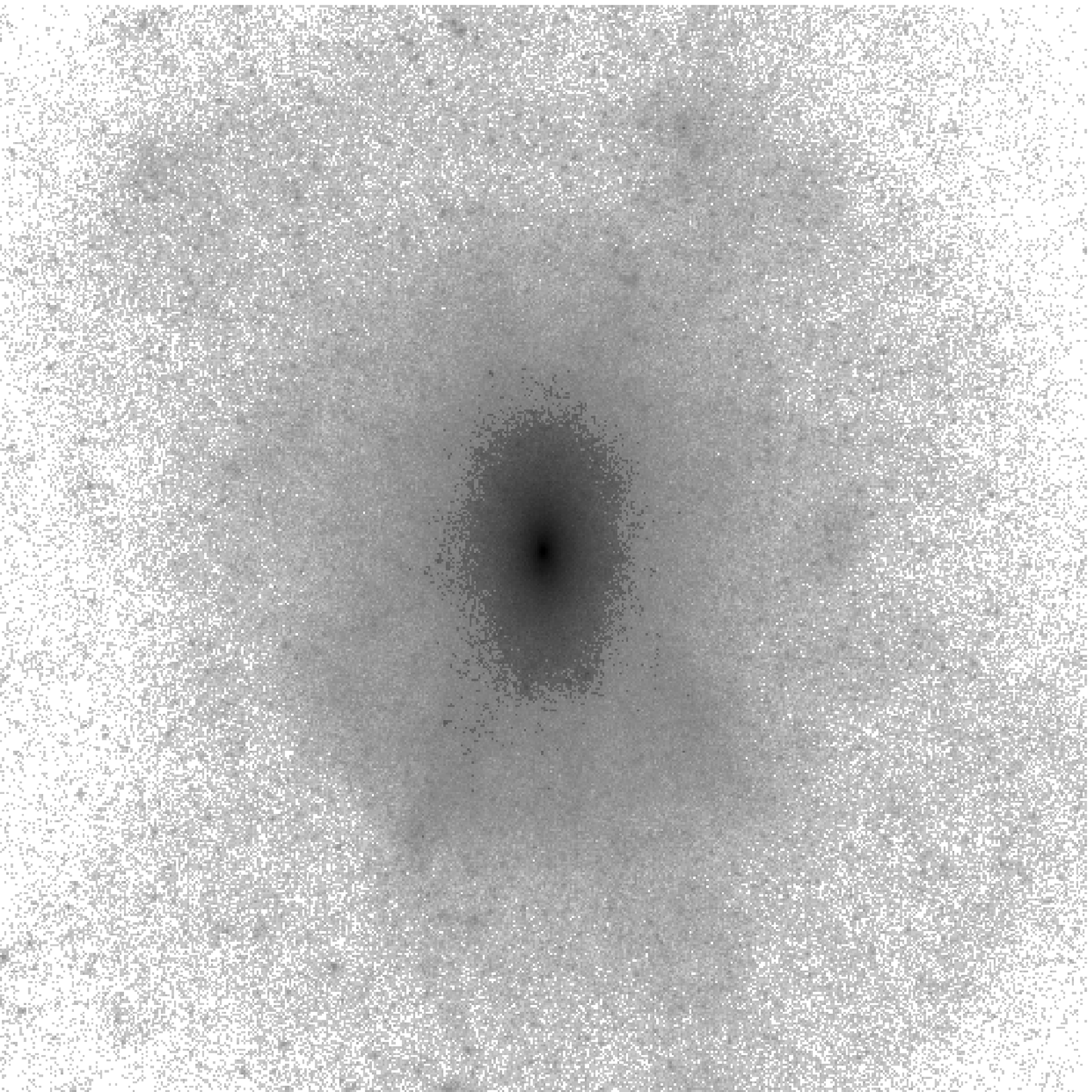} 
\caption{The largest halo in the model with standard spectrum D0 (\textit{left}) and with dip 16D1000 (\textit{right}). Each dot represents individual probe particle. The darker color means higher local density.}
\label{fig:haloes}
\end{figure}

Another important consideration is abundance of subhaloes within the haloes. It is evident that lack of power on certain scales in spectrum of perturbations results in significant quantity reduction of corresponding mass haloes. We did not explore this question quantitatively. However, the effect can be easily seen in fig.~\ref{fig:haloes} where the halo in standard model is inhabited by numerous satellites, whereas this halo in featured model has no large subhaloes at all. This fact imposes restrictions on number and masses of satellites of cuspless galaxies. Specifically, such galaxies must not have large satellites. Furthermore, one can expect that the largest subhaloes in the featured models can form from the perturbations which correspond to the right edge of the dip. Given that the right edge scale ($k_2$) is approximately one order of magnitude smaller than that of the left edge ($k_1$), the largest subhaloes mass can be about $10^{-3}$ of that of the main halo. Since the results are scale free, for dwarf galaxies (where cores are observed) the largest mass of subhaloes can be estimated as $\sim 10^7$ \hM. This allowed mass of satellites is too small to start star formation. For the sake of further investigation of this problem it would be interesting to perform large cosmological simulation with modified spectrum to investigate the properties of haloes' satellites in the same fashion like it was done by \citet{kravtsov_msp} for standard spectrum.

\section{Conclusions}

The inner structure of dark matter haloes was studied in models with different spectral features. It was found that having low amplitudes of intermediate perturbations (i.e. modes with scales from the main scale to 8--32 times smaller one), density cusps flattens from initial $\alpha \approx -1.2$ to $\alpha \approx -0.6$ or even to $\alpha \approx -0.2$. Partial or entire cusp reproduction in case if all small-scales fluctuations are suppressed indicates that it is the small-scale modes which are responsible for the cusp flattening. According to the entropy theory these small-scale perturbations converts into fast random motions of dark matter particles which destruct the central density cusp. Investigation of features variety enables us to define optimal parameters of intermediate modes suppression to destroy density cusp. In particular, we found that:
\begin{itemize}
\item The effect occurs if intermediate modes are suppressed by more than 10 times. The higher rate of suppression -- the more pronounced effect is observed.

\item To eliminate the cusp intermediate perturbations must be suppressed on a range from the main halo scale to 8--32 times smaller one. In the model with narrow dip (4D1000) cusp reproduces due to intermediate modes (with wave numbers $4k_h < k < 8k_h$) influence, while in the model with too broad dip (CUT1000) the cusp can not be destroyed since there all small-scale fluctuations are suppressed.

\item The effect occurs only if the halo's scale lies on the left boundary of the dip. Partial suppression of the main scale results in significant drop of central mass.

\item Pure enhancing of small-scale perturbations cannot destroy the cusp. 

\item The scatter of central slopes of haloes with similar masses is quite large and may be comparable with the effect itself.
\end{itemize}

There are two possible ways of how this effect of cusp flattening can act in our Universe. First one was suggested by \citet{PDLM:2012}: actual power spectrum bears purely statistical value and some of its local realisations in the standard model can be similar to the studied modified spectra. \citet{PDLM:2012} found the probability of such statistical suppression of intermediate modes to be about $10^{-2}$. Taken account of relative abundance of special types of galaxies where cores are observed, this explanation seems plausible. In these terms the core-cusp problem results from selection bias. However, detailed investigation of this purely statistical effect in N-body simulations is beyond the capability of state-of-art numerical cosmology since it requires a large (about $10^3$) sample of well resolved haloes (i.e. haloes which contain at least $10^6$ particles each).

We suggest here the second and more promising way. If the dark matter consists of more than one species (at least two are needed), it can naturally produce global initial power spectrum with a dip. In this model the large scale part of the spectrum is provided by lightweight warm particles with mass 10-30 keV. The density fluctuations on the scales smaller than that of dwarf or LSB galaxies with observed cores are damped. This damping creates the left edge of the spectral dip. At smaller scales the other, heavier component creates base of the dip with amplitude at least 10 times lower than that of the standard CDM spectrum. The \textit{isocurvature} perturbations can produce the right edge of the dip at smaller scales.

Nowadays, $\Lambda$CDM power spectrum is accurately measured to minimal scales of about 1 \hMpc. Moreover, nowadays we have evidence of small-scale perturbations power shortage on scales of \mbox{10--300 \hkpc}  \citep{2003ApJ...597...81D}. So the complex power spectrum on small scales is not forbidden by current observations. One of predictions of our model is the deficit of small haloes on some scale. This deficit can be manifested as the satellite abundance problem. Simulations with larger volume and observations of satellites of galaxies should be used to judge this model.

\bibliography{biblio}

\begin{thebibliography}{}

\bibitem[\protect\citeauthoryear{{Abdelqader} \& {Melia}}{{Abdelqader} \&
  {Melia}}{2008}]{decdm1}
{Abdelqader} M.,  {Melia} F.,  2008, \mnras, 388, 1869

\bibitem[\protect\citeauthoryear{{Antonov}}{{Antonov}}{1961}]{antonov}
{Antonov} V.~A.,  1961, \sovast, 4, 859

\bibitem[\protect\citeauthoryear{{Arabadjis}, {Bautz} \& {Garmire}}{{Arabadjis}
  et~al.}{2002}]{2002ApJ...572...66A}
{Arabadjis} J.~S.,  {Bautz} M.~W.,    {Garmire} G.~P.,  2002, \apj, 572, 66

\bibitem[\protect\citeauthoryear{{Burkert}}{{Burkert}}{1995}]{Burkert1995}
{Burkert} A.,  1995, \apjl, 447, L25

\bibitem[\protect\citeauthoryear{{Cen}}{{Cen}}{2001}]{decdm0}
{Cen} R.,  2001, \apjl, 546, L77

\bibitem[\protect\citeauthoryear{{Col{\'{\i}}n}, {Avila-Reese} \&
  {Valenzuela}}{{Col{\'{\i}}n} et~al.}{2000}]{wdm}
{Col{\'{\i}}n} P.,  {Avila-Reese} V.,    {Valenzuela} O.,  2000, \apj, 542, 622

\bibitem[\protect\citeauthoryear{{de Blok}}{{de Blok}}{2010}]{deBlok2010}
{de Blok} W.~J.~G.,  2010, Advances in Astronomy, 2010

\bibitem[\protect\citeauthoryear{{de Blok}, {McGaugh}, {Bosma} \& {Rubin}}{{de
  Blok} et~al.}{2001}]{deblok}
{de Blok} W.~J.~G.,  {McGaugh} S.~S.,  {Bosma} A.,    {Rubin} V.~C.,  2001,
  \apjl, 552, L23

\bibitem[\protect\citeauthoryear{{Demia{\'n}ski} \&
  {Doroshkevich}}{{Demia{\'n}ski} \&
  {Doroshkevich}}{2003}]{2003ApJ...597...81D}
{Demia{\'n}ski} M.,  {Doroshkevich} A.,  2003, \apj, 597, 81

\bibitem[\protect\citeauthoryear{Doroshkevich, Lukash \& Mikheeva}{Doroshkevich
  et~al.}{2012}]{DoLuMa}
Doroshkevich A.~G.,  Lukash V.~N.,    Mikheeva E.~V.,  2012, \ufnE, 182, 3

\bibitem[\protect\citeauthoryear{{Ho} \& {Scherrer}}{{Ho} \& {Scherrer}}{2013}]{anapole}
{Ho} C.~M., {Scherrer} R.~J., 2013, Physics Letters B, 722, 341

\bibitem[\protect\citeauthoryear{{Knebe}, {Islam} \& {Silk}}{{Knebe}
  et~al.}{2001}]{2001MNRAS.326..109K}
{Knebe} A.,  {Islam} R.~R.,    {Silk} J.,  2001, \mnras, 326, 109

\bibitem[\protect\citeauthoryear{{Kravtsov}, {Gnedin} \& {Klypin}}{{Kravtsov}
  et~al.}{2004}]{kravtsov_msp}
{Kravtsov} A.~V.,  {Gnedin} O.~Y.,    {Klypin} A.~A.,  2004, \apj, 609, 482

\bibitem[\protect\citeauthoryear{{Kravtsov}, {Klypin}, {Bullock} \&
  {Primack}}{{Kravtsov} et~al.}{1998}]{kravtsov1998}
{Kravtsov} A.~V.,  {Klypin} A.~A.,  {Bullock} J.~S.,    {Primack} J.~R.,  1998,
  \apj, 502, 48

\bibitem[\protect\citeauthoryear{{Lewis}, {Buote} \& {Stocke}}{{Lewis}
  et~al.}{2003}]{2003ApJ...586..135L}
{Lewis} A.~D.,  {Buote} D.~A.,    {Stocke} J.~T.,  2003, \apj, 586, 135

\bibitem[\protect\citeauthoryear{{Lynden-Bell}}{{Lynden-Bell}}{1967}]{lyndenbell}
{Lynden-Bell} D.,  1967, \mnras, 136, 101

\bibitem[\protect\citeauthoryear{{Martizzi}, {Teyssier} \& {Moore}}{{Martizzi}
  et~al.}{2013}]{2013MNRAS.tmp.1290M}
{Martizzi} D.,  {Teyssier} R.,    {Moore} B.,  2013, \mnras

\bibitem[\protect\citeauthoryear{{Medvedev}}{{Medvedev}}{2012}]{2012APS..APR.G7007M}
{Medvedev} M.~V.,  2012, in APS April Meeting Abstracts {Cosmological
  Simulations Evidence in Favor of Two-Component Flavor-Mixed Cold Dark
  Matter}.
p. G7007

\bibitem[\protect\citeauthoryear{{Peebles}}{{Peebles}}{2000}]{fluiddm}
{Peebles} P.~J.~E.,  2000, \apjl, 534, L127

\bibitem[\protect\citeauthoryear{{Pilipenko}, {Doroshkevich} \&
  {Gottl{\"o}ber}}{{Pilipenko} et~al.}{2009}]{2009ARep...53..976P}
{Pilipenko} S.~V.,  {Doroshkevich} A.~G.,    {Gottl{\"o}ber} S.,  2009,
  Astronomy Reports, 53, 976

\bibitem[\protect\citeauthoryear{{Pilipenko}, {Doroshkevich}, {Lukash} \&
  {Mikheeva}}{{Pilipenko} et~al.}{2012}]{PDLM:2012}
{Pilipenko} S.~V.,  {Doroshkevich} A.~G.,  {Lukash} V.~N.,    {Mikheeva} E.~V.,
   2012, \mnras, 427, L30

\bibitem[\protect\citeauthoryear{{Salucci} \& {Burkert}}{{Salucci} \&
  {Burkert}}{2000}]{salucci2000}
{Salucci} P.,  {Burkert} A.,  2000, \apjl, 537, L9

\bibitem[\protect\citeauthoryear{{Shapiro}, {Iliev}, {Martel}, {Ahn} \&
  {Alvarez}}{{Shapiro} et~al.}{2004}]{2004astro.ph..9173S}
{Shapiro} P.~R.,  {Iliev} I.~T.,  {Martel} H.,  {Ahn} K.,    {Alvarez} M.~A.,
  2004, ArXiv Astrophysics e-prints

\bibitem[\protect\citeauthoryear{{Spergel} \& {Steinhardt}}{{Spergel} \&
  {Steinhardt}}{2000}]{sidm}
{Spergel} D.~N.,  {Steinhardt} P.~J.,  2000, Physical Review Letters, 84, 3760

\bibitem[\protect\citeauthoryear{{Springel}}{{Springel}}{2005}]{gadget2}
{Springel} V.,  2005, \mnras, 364, 1105

\bibitem[\protect\citeauthoryear{{Tremaine} \& {Gunn}}{{Tremaine} \&
  {Gunn}}{1979}]{wdm0}
{Tremaine} S.,  {Gunn} J.~E.,  1979, Physical Review Letters, 42, 407

\bibitem[\protect\citeauthoryear{{Zhao}}{{Zhao}}{1996}]{zhao1996}
{Zhao} H.,  1996, \mnras, 278, 488

\end{thebibliography}

\begin{table*}
\caption{The inner slope and mass deficiency within $r_s$ (with respect to the model with standard spectrum) in investigated models. Each column corresponds to one of the haloes (H1--H5), which parameters (virial mass and concentration in the standard model) are given at the top of the table. The first lines in each row represent central slopes $\alpha$ and the second lines stand for the mass shortages. The models are split into sets in the same way as in fig.~\ref{fig:models}.}
\label{tab:results}
\begin{tabular}{l c c c c c }
\hline
\hline
      & \textbf{H1} & \textbf{H2} & \textbf{H3} & \textbf{H4} & \textbf{H5} \\
\hline
$M_{vir}$, $10^{13}$ \hM & 9.9 & 4.8 & 2.4 & 2.0 & 1.7 \\
$c\equiv R_{vir}/r_s$ & 8.9 & 13.0 & 13.2 & 10.2 & 12.9 \\
\hline
\hline
\textbf{Model} & \multicolumn{5}{c}{\textbf{Central slope} $\alpha$} \\
               & \multicolumn{5}{c}{\textbf{Central mass deficiency} $(M_{\rm D0}(r_s)-M(r_s))/M_{\rm D0}(r_s)$, \%} \\
\hline
&\multicolumn{5}{c}{ \textit{Standard spectrum:} } \\[1mm]
%\hline
D0 & $-1.23 \pm 0.03$ & $-1.20 \pm 0.10$ & $-1.35 \pm 0.05$ & $-1.24 \pm 0.04$ & $-1.30 \pm 0.05$ \\
 &0\% & 0\% & 0\% & 0\% & 0\% \\[2mm]
\hline
&\multicolumn{5}{c}{ \textit{Various depth:} } \\[1mm]
%\hline
16D10 & $-0.98 \pm 0.02$ & $-1.10 \pm 0.09$ & $-1.20 \pm 0.19$ & $-0.71 \pm 0.02$ & $-0.88 \pm 0.19$\\ 
 &10\% & 20\% & 30\% & 44\% & 41\% \\[2mm]
16D100 & $-0.90 \pm 0.04$ & $-0.84 \pm 0.09$ & $-1.02 \pm 0.10$ & $-0.67 \pm 0.19$ & $-0.99 \pm 0.04$\\
 &14\% & 29\% & 42\% & 53\% & 58\% \\[2mm]
16D1000 & $-0.82 \pm 0.04$ & $-0.73 \pm 0.14$ & $-0.94 \pm 0.06$ & $-0.60 \pm 0.25$ & $-0.70 \pm 0.08$\\ 
 &17\% & 31\% & 43\% & 52\% & 58\% \\[2mm]
\hline
&\multicolumn{5}{c}{ \textit{Various width:} } \\[2mm]
%\hline
4D1000 & $-0.84 \pm 0.02$ & $-1.21 \pm 0.05$ & $-1.21 \pm 0.16$ & $-1.31 \pm 0.22$ & $-1.33 \pm 0.10$ \\
&16\% & 34\% & 42\% & 47\% & 63\% \\[2mm]
8D1000 & $-0.70 \pm 0.12$ & $-0.92 \pm 0.03$ & $-0.93 \pm 0.05$ & $-0.20 \pm 0.14$ & $-0.94 \pm 0.11$\\
&12\% & 33\% & 42\% & 59\% & 61\% \\[2mm] 
16D1000 & $-0.82 \pm 0.04$ & $-0.73 \pm 0.14$ & $-0.94 \pm 0.06$ & $-0.60 \pm 0.25$ & $-0.70 \pm 0.08$\\ 
&17\% & 31\% & 43\% & 52\% & 58\% \\[2mm]
32D1000 & $-0.87 \pm 0.11$ & $-1.03 \pm 0.03$ & $-0.90 \pm 0.06$ & $-0.54 \pm 0.09$ & $-0.68 \pm 0.13$\\ 
&15\% & 29\% & 40\% & 58\% & 51\% \\[2mm]
CUT1000 & $-1.12 \pm 0.11$ & $-0.93 \pm 0.13$ & $-1.03 \pm 0.21$ & $-0.71 \pm 0.17$ & $-0.85 \pm 0.19$\\ 
&14\% & 27\% & 40\% & 56\% & 51\% \\[2mm]
\hline
&\multicolumn{5}{c}{ \textit{Various position:} } \\[2mm]
%\hline
8D1000 & $-0.70 \pm 0.12$ & $-0.92 \pm 0.03$ & $-0.93 \pm 0.05$ & $-0.20 \pm 0.14$ & $-0.94 \pm 0.11$\\ 
&12\% & 33\% & 42\% & 59\% & 61\% \\[2mm]
8D1000\_HF & $-0.93 \pm 0.06$ & $-1.40 \pm 0.03$ & $-1.37 \pm 0.04$ & $-1.29 \pm 0.08$ & $-1.02 \pm 0.10$ \\
&10\% & 13\% & 17\% & 12\% & 5\% \\[2mm] 
\hline
&\multicolumn{5}{c}{ \textit{Bumpy spectra:} } \\[2mm]
%\hline
16D1000B200 & $-0.53 \pm 0.04$ & $-0.75 \pm 0.18$ & $-1.10 \pm 0.11$ & $-1.03 \pm 0.28$ & $-1.22 \pm 0.15$\\ 
&12\% & 28\% & 45\% & 55\% & 64\% \\[2mm]
16B200 & $-1.18 \pm 0.10$ & $-1.19 \pm 0.10$ & $-1.32 \pm 0.05$ & $-1.37 \pm 0.09$ & $-1.13 \pm 0.04$\\
&$-7$\% & 11\% & 2\% & $-1$\% & 0\% \\[2mm]
\hline
\hline
\end{tabular}
\end{table*}

\end{document}